\documentclass[10pt,letterpaper]{article}
\usepackage{opex3}

\usepackage{graphicx}
\usepackage{epstopdf}

\usepackage{cite}

\usepackage{amsmath}

\usepackage{amsfonts}

\begin{document}

\title{Interactions of Airy beams, nonlinear accelerating beams, and induced solitons in Kerr and saturable nonlinear media}

\author{Yiqi Zhang,$^{1,4}$ Milivoj R. Beli\'c,$^{2,\ast}$ Huaibin Zheng,$^1$ Haixia Chen,$^1$ Changbiao Li,$^1$ Yuanyuan Li,$^3$ and Yanpeng Zhang$^{1,5}$}
\address{%
 $^1$Key Laboratory for Physical Electronics and Devices of the Ministry of Education \& Shaanxi Key Lab of Information Photonic Technique,
Xi'an Jiaotong University, Xi'an 710049, China \\
$^2$Science Program, Texas A\&M University at Qatar, P.O. Box 23874 Doha, Qatar\\
$^3$Institute of Applied Physics, Xi'an University of Arts and Science, Xi'an 710065, China \\
$^4$zhangyiqi@mail.xjtu.edu.cn; http://zhangyiqi.gr.xjtu.edu.cn \\
$^5$ypzhang@mail.xjtu.edu.cn; http://ypzhang.gr.xjtu.edu.cn\\
\email{$^*$Corresponding author: milivoj.belic@qatar.tamu.edu }
}


\begin{abstract}
  We investigate numerically interactions between two in-phase or out-of-phase
  Airy beams and nonlinear accelerating beams in Kerr
  and saturable nonlinear media in one transverse dimension.
  We discuss different cases in which the beams with different intensities are
  launched into the medium, but accelerate in opposite directions.
  Since both the Airy beams and nonlinear accelerating beams possess infinite oscillating tails,
  we discuss interactions between truncated beams, with finite energies.
  During interactions we see solitons and soliton pairs generated that are not accelerating.
  In general, the higher the intensities of interacting beams, the easier to form solitons;
  when the intensities are small enough, no solitons are generated.
  Upon adjusting the interval between the launched beams,
  their interaction exhibits different properties.
  If the interval is large relative to the width of the first lobes,
  the generated soliton pairs just propagate individually and do not interact much.
  However, if the interval is comparable to the widths of the maximum lobes,
  the pairs strongly interact and display varied behavior.
\end{abstract}



\bibliography{refs_airy}

\begin{thebibliography}{10}
\newcommand{\enquote}[1]{``#1''}

\bibitem{siviloglou_ol_2007}
G.~A. Siviloglou and D.~N. Christodoulides, \enquote{Accelerating finite energy
  {A}iry beams,} Opt. Lett. \textbf{32}, 979--981 (2007).

\bibitem{siviloglou_prl_2007}
G.~A. Siviloglou, J.~Broky, A.~Dogariu, and D.~N. Christodoulides,
  \enquote{Observation of accelerating {Airy} beams,} Phys. Rev. Lett.
  \textbf{99}, 213901 (2007).

\bibitem{bandres_ol_2008}
M.~A. Bandres, \enquote{Accelerating parabolic beams,} Opt. Lett. \textbf{33},
  1678--1680 (2008).

\bibitem{bandres_ol_2009}
M.~A. Bandres, \enquote{Accelerating beams,} Opt. Lett. \textbf{34}, 3791--3793
  (2009).

\bibitem{ellenbogen_np_2009}
T.~Ellenbogen, N.~Voloch-Bloch, A.~Ganany-Padowicz, and A.~Arie,
  \enquote{Nonlinear generation and manipulation of {A}iry beams,} Nat. Photon.
  \textbf{3}, 395--398 (2009).

\bibitem{chong_np_2010}
A.~Chong, W.~H. Renninger, D.~N. Christodoulides, and F.~W. Wise,
  \enquote{{A}iry-{B}essel wave packets as versatile linear light bullets,}
  Nat. Photon. \textbf{4}, 103--106 (2010).

\bibitem{efremidis_ol_2010}
N.~K. Efremidis and D.~N. Christodoulides, \enquote{Abruptly autofocusing
  waves,} Opt. Lett. \textbf{35}, 4045--4047 (2010).

\bibitem{alonso_ol_2012}
M.~A. Alonso and M.~A. Bandres, \enquote{Spherical fields as nonparaxial
  accelerating waves,} Opt. Lett. \textbf{37}, 5175--5177 (2012).

\bibitem{kaminer_oe_2012}
I.~Kaminer, J.~Nemirovsky, and M.~Segev, \enquote{Self-accelerating
  self-trapped nonlinear beams of {M}axwell's equations,} Opt. Express
  \textbf{20}, 18827--18835 (2012).

\bibitem{kaminer_prl_2012}
I.~Kaminer, R.~Bekenstein, J.~Nemirovsky, and M.~Segev, \enquote{Nondiffracting
  accelerating wave packets of {M}axwell's equations,} Phys. Rev. Lett.
  \textbf{108}, 163901 (2012).

\bibitem{aleahmad_prl_2012}
P.~Aleahmad, M.-A. Miri, M.~S. Mills, I.~Kaminer, M.~Segev, and D.~N.
  Christodoulides, \enquote{Fully vectorial accelerating diffraction-free
  {H}elmholtz beams,} Phys. Rev. Lett. \textbf{109}, 203902 (2012).

\bibitem{bandres_oe_2013}
M.~A. Bandres, M.~A. Alonso, I.~Kaminer, and M.~Segev,
  \enquote{Three-dimensional accelerating electromagnetic waves,} Opt. Express
  \textbf{21}, 13917--13929 (2013).

\bibitem{bandres_njp_2013}
M.~A. Bandres and B.~M. Rodr\'{\i}guez-Lara, \enquote{Nondiffracting
  accelerating waves: {W}eber waves and parabolic momentum,} New J. Phys.
  \textbf{15}, 013054 (2013).

\bibitem{hu_ol_2010}
Y.~Hu, S.~Huang, P.~Zhang, C.~Lou, J.~Xu, and Z.~Chen, \enquote{Persistence and
  breakdown of {A}iry beams driven by an initial nonlinearity,} Opt. Lett.
  \textbf{35}, 3952--3954 (2010).

\bibitem{kaminer_prl_2011}
I.~Kaminer, M.~Segev, and D.~N. Christodoulides, \enquote{Self-accelerating
  self-trapped optical beams,} Phys. Rev. Lett. \textbf{106}, 213903 (2011).

\bibitem{efremidis_pra_2013}
N.~K. Efremidis, V.~Paltoglou, and W.~von Klitzing, \enquote{Accelerating and
  abruptly autofocusing matter waves,} Phys. Rev. A \textbf{87}, 043637 (2013).

\bibitem{salandrino_ol_2010}
A.~Salandrino and D.~N. Christodoulides, \enquote{Airy plasmon: a
  nondiffracting surface wave,} Opt. Lett. \textbf{35}, 2082--2084 (2010).

\bibitem{zhang_ol_2011b}
P.~Zhang, S.~Wang, Y.~Liu, X.~Yin, C.~Lu, Z.~Chen, and X.~Zhang,
  \enquote{Plasmonic {A}iry beams with dynamically controlled trajectories,}
  Opt. Lett. \textbf{36}, 3191--3193 (2011).

\bibitem{minovich_prl_2011}
A.~Minovich, A.~E. Klein, N.~Janunts, T.~Pertsch, D.~N. Neshev, and Y.~S.
  Kivshar, \enquote{Generation and near-field imaging of {A}iry surface
  plasmons,} Phys. Rev. Lett. \textbf{107}, 116802 (2011).

\bibitem{li_prl_2011}
L.~Li, T.~Li, S.~M. Wang, C.~Zhang, and S.~N. Zhu, \enquote{Plasmonic {A}iry
  beam generated by in-plane diffraction,} Phys. Rev. Lett. \textbf{107},
  126804 (2011).

\bibitem{li_nano_2011}
L.~Li, T.~Li, S.~Wang, S.~Zhu, and X.~Zhang, \enquote{Broad band focusing and
  demultiplexing of in-plane propagating surface plasmons,} Nano Lett.
  \textbf{11}, 4357--4361 (2011).

\bibitem{zhuang_ol_2012b}
F.~Zhuang, J.~Shen, X.~Du, and D.~Zhao, \enquote{Propagation and modulation of
  {Airy} beams through a four-level electromagnetic induced transparency atomic
  vapor,} Opt. Lett. \textbf{37}, 3054--3056 (2012).

\bibitem{zhuang_ol_2012a}
F.~Zhuang, X.~Du, Y.~Ye, and D.~Zhao, \enquote{Evolution of {A}iry beams in a
  chiral medium,} Opt. Lett. \textbf{37}, 1871--1873 (2012).

\bibitem{kaminer_oe_2013}
I.~Kaminer, J.~Nemirovsky, K.~G. Makris, and M.~Segev,
  \enquote{Self-accelerating beams in photonic crystals,} Opt. Express
  \textbf{21}, 8886--8896 (2013).

\bibitem{durnin_josaa_1987}
J.~Durnin, \enquote{Exact solutions for nondiffracting beams. {I}. the scalar
  theory,} J. Opt. Soc. Am. A \textbf{4}, 651--654 (1987).

\bibitem{zhang_ol_2012}
P.~Zhang, Y.~Hu, D.~Cannan, A.~Salandrino, T.~Li, R.~Morandotti, X.~Zhang, and
  Z.~Chen, \enquote{Generation of linear and nonlinear nonparaxial accelerating
  beams,} Opt. Lett. \textbf{37}, 2820--2822 (2012).

\bibitem{zhang_prl_2012}
P.~Zhang, Y.~Hu, T.~Li, D.~Cannan, X.~Yin, R.~Morandotti, Z.~Chen, and
  X.~Zhang, \enquote{Nonparaxial {Mathieu} and {Weber} accelerating beams,}
  Phys. Rev. Lett. \textbf{109}, 193901 (2012).

\bibitem{liu_ol_2013}
S.~Liu, M.~Wang, P.~Li, P.~Zhang, and J.~Zhao, \enquote{Abrupt polarization
  transition of vector autofocusing {A}iry beams,} Opt. Lett. \textbf{38},
  2416--2418 (2013).

\bibitem{ament_prl_2011}
C.~Ament, P.~Polynkin, and J.~V. Moloney, \enquote{Supercontinuum generation
  with femtosecond self-healing airy pulses,} Phys. Rev. Lett. \textbf{107},
  243901 (2011).

\bibitem{dolev_prl_2012}
I.~Dolev, I.~Kaminer, A.~Shapira, M.~Segev, and A.~Arie, \enquote{Experimental
  observation of self-accelerating beams in quadratic nonlinear media,} Phys.
  Rev. Lett. \textbf{108}, 113903 (2012).

\bibitem{fattal_oe_2011}
Y.~Fattal, A.~Rudnick, and D.~M. Marom, \enquote{Soliton shedding from {A}iry
  pulses in {K}err media,} Opt. Express \textbf{19}, 17298--17307 (2011).

\bibitem{driben_pra_2013}
R.~Driben, B.~A. Malomed, A.~V. Yulin, and D.~V. Skryabin, \enquote{Newton's
  cradles in optics: From $\mathcal{N}$-soliton fission to soliton chains,}
  Phys. Rev. A \textbf{87}, 063808 (2013).

\bibitem{zhang_ol_2013}
Y.~Q. Zhang, M.~Beli\'{c}, Z.~K. Wu, H.~B. Zheng, K.~Q. Lu, Y.~Y. Li, and Y.~P.
  Zhang, \enquote{Soliton pair generation in the interactions of airy and
  nonlinear accelerating beams,} Opt. Lett. \textbf{38}, 4585--4588 (2013).

\bibitem{berry_ajp_1979}
M.~V. Berry and N.~L. Balazs, \enquote{Nonspreading wave packets,} Am. J. Phys.
  \textbf{47}, 264--267 (1979).

\bibitem{yang_book}
J.~Yang, \emph{Nonlinear waves in integrable and non-integrable systems} (SIAM,
  Philadelphia, 2010).

\end{thebibliography}
\bibliographystyle{osajnl}


\section{Introduction}
Self-accelerating nondiffracting optical beams are special wavepackets that exhibit self-accelerating,
nondiffracting, and self-healing properties during propagation \cite{siviloglou_ol_2007,siviloglou_prl_2007}.
In the last decade such optical beams have been extensively studied
\cite{bandres_ol_2008, bandres_ol_2009, ellenbogen_np_2009,chong_np_2010,efremidis_ol_2010,alonso_ol_2012,
kaminer_oe_2012,kaminer_prl_2012,aleahmad_prl_2012,bandres_oe_2013,bandres_njp_2013}, mostly in optically linear media.
In the past few years, the involved media also included nonlinear dielectrics \cite{hu_ol_2010,kaminer_prl_2011},
Bose-Einstein condensates \cite{efremidis_pra_2013},
the surface of a metal \cite{salandrino_ol_2010,zhang_ol_2011b,minovich_prl_2011,li_prl_2011,li_nano_2011},
atomic vapors with electromagnetically induced transparency \cite{zhuang_ol_2012b},
chiral media \cite{zhuang_ol_2012a},
photonic crystals \cite{kaminer_oe_2013}, and so on.
All along, special attention has been focused on Airy \cite{siviloglou_ol_2007,siviloglou_prl_2007}
and Bessel beams \cite{durnin_josaa_1987}.
Analyses have been confined to linear media, for the reason of wanting to observe minimally diffracting beams in linear optics.
According to the linear Schr\"{o}dinger equation (SE), the beam or the wave packet in the form of Airy function evolves
practically without diffraction and accelerates along a parabolic trajectory
\cite{siviloglou_ol_2007,siviloglou_prl_2007}. In addition to the paraxial accelerating beams,
nonparaxial beams have also attracted a lot of attention \cite{kaminer_oe_2012,kaminer_prl_2012,zhang_ol_2012, aleahmad_prl_2012, zhang_prl_2012},
their analysis being based on Maxwell's equations.

Thus far single accelerating beams have been thoroughly investigated, including their dynamics and properties in propagation.
Compared to that, interactions between Airy beams have attracted little attention.
Although radially symmetric Airy beams display self-focusing in a nonlinear (NL) medium
\cite{efremidis_ol_2010,efremidis_pra_2013,liu_ol_2013},
the interaction between two Airy beams with varying distance between them has not been studied much.
Until now, the experimental observation of Airy beams \cite{siviloglou_prl_2007},
Airy beams in NL materials \cite{ellenbogen_np_2009,hu_ol_2010},
and accelerating nonlinear beams \cite{ament_prl_2011,zhang_ol_2012,dolev_prl_2012}
in NL materials has been reported.
In theory, the splitting of Airy waves into solitons in a Kerr medium was analyzed in \cite{fattal_oe_2011}
and related to this, the splitting of higher-order solitons into a chain of fundamental solitons
under the action of third-order dispersion was reported in \cite{driben_pra_2013}.
But, the interaction of two Airy beams or two nonlinear accelerating beams is barely discussed.
In a recent paper we have investigated the interactions of Airy beams and accelerating nonlinear beams
in a Kerr and a saturable photorefractive medium \cite{zhang_ol_2013}.
Still, many interesting questions remain unanswered, such as:
Can accelerating beams emerge from the interactions?
Are these interactions elastic?
How about interactions between accelerating beams and solitons?
In this paper we try to answer these questions through a systematic study of the interactions of Airy beams
and nonlinear accelerating beams in Kerr and saturable media.

The organization of the paper is as follows.
We briefly introduce the theoretical model in Sec. \ref{theory};
in Sec. \ref{airyairy}, we investigate the interactions of two Airy beams;
in Sec. \ref{nonnon} we discuss the interactions of nonlinear accelerating beams; and
in Sec. \ref{different} we study the interactions of different accelerating beams in more detail.
Section \ref{conclusion} concludes the paper.

\section{Basic theory}\label{theory}
The scaled equation for the propagation of a slowly-varying envelope $\psi$ of the optical electric field
in one transverse dimension in the paraxial approximation is of the NLSE form:

\begin{equation}
\label{eq1}
  i\frac{\partial \psi}{\partial z} +\frac{1}{2} \frac{\partial^2 \psi}{\partial x^2} +\delta n \psi=0,
\end{equation}
where $\delta n$ -- a function of the intensity $|\psi(x)|^2$ in a NL medium -- represents the change in the index of refraction, and
$x$ and $z$ are the dimensionless transverse and the propagation coordinates, respectively. They are
measured in units of some typical transverse size $x_0$ and the corresponding Rayleigh range $kx_0^2$.
Without $\delta n$ in Eq. (\ref{eq1}), the equation is just the linear SE, allowing many particular solutions.
One of the well-known exact solutions is the Airy wave \cite{berry_ajp_1979}
\begin{equation*}
  \psi(x,z)={\rm Ai}\left( x-\frac{z^2}{4} \right) \exp \left[ i \left( \frac{xz}{2}-\frac{z^3}{12} \right) \right],
\end{equation*}
with a characteristic infinite oscillatory tail. The tail makes the wave of infinite energy. To make it finite-energy,
the solution is generalized into \cite{siviloglou_ol_2007}
\begin{align}
\label{eq2} \psi(x,z)= {\rm Ai}\left( x-\frac{z^2}{4}+iaz \right)
\exp \left[ \frac{i}{12}(6a^2z-12iax +6iaz^2
+6xz-z^3)\right],
\end{align}
with an arbitrary real decay constant $a\ge0$.
This solution is generated from an initial condition $  {\psi(x)={\rm Ai} (x) \exp(ax)}$
and represents a finite-energy Airy wave. This wave was the first of the peculiar solutions to
the linear SE, observed to display transverse self-acceleration \cite{berry_ajp_1979}.

More interesting accelerating solutions to the NLSE -- the nonlinear accelerating beams -- are constructed
from Eq. (\ref{eq1}), by introducing a traveling variable $x-z^2/4$ to substitute for $x$
\cite{kaminer_prl_2011}:
\begin{equation}
\label{eq3}
  i\frac{\partial \psi}{\partial z} - i\frac{z}{2} \frac{\partial \psi}{\partial x} +\frac{1}{2}
  \frac{\partial^2 \psi}{\partial x^2} +\delta n \psi=0.
\end{equation}
We seek NL self-trapped solutions of Eq. (\ref{eq3}) in the
form $\psi(x,z)=u(x)\exp{[i(xz/2+z^3/24)]}$ that accelerate along a parabolic trajectory;
this leads to the equation
\begin{equation}
\label{eq4}
  \frac{\partial^2 u}{\partial x^2} +2\delta n u -x u=0.
\end{equation}
We treat Eq. (\ref{eq4}) as an initial value problem with the asymptotic behavior
$u(x)=\alpha\textrm{Ai}(x)$ and $u'(x)=\alpha\textrm{Ai}'(x)$ for large enough $x>0$;
here $\alpha$ indicates the strength of the nonlinearity induced by the potential solution.

In this paper the nonlinearity of the medium is assumed to be of the Kerr type or of the saturable type only,
so that the nonlinearity is $\delta n \propto |\psi|^2$ or $\delta n \propto |\psi|^2/(r+|\psi|^2)$;
these are even functions of the transverse coordinate $x$.
As a result, the solution of Eq. (\ref{eq1}) with, as well as without, $\delta n$ will not be affected if it
is reversed about $x=0$ and shifted along the transverse coordinate.
Here $0 < r \leq 1$ is the saturation parameter, for which, without loss of generality, we assume $r=1$.
The saturation parameter is connected with the dark current or the background illumination in the
medium, so that the strength of the nonlinearity becomes smaller with bigger $r$, but the interaction remains quite similar.

\begin{figure}[htbp]
  \centering
  \includegraphics[width=0.7\columnwidth]{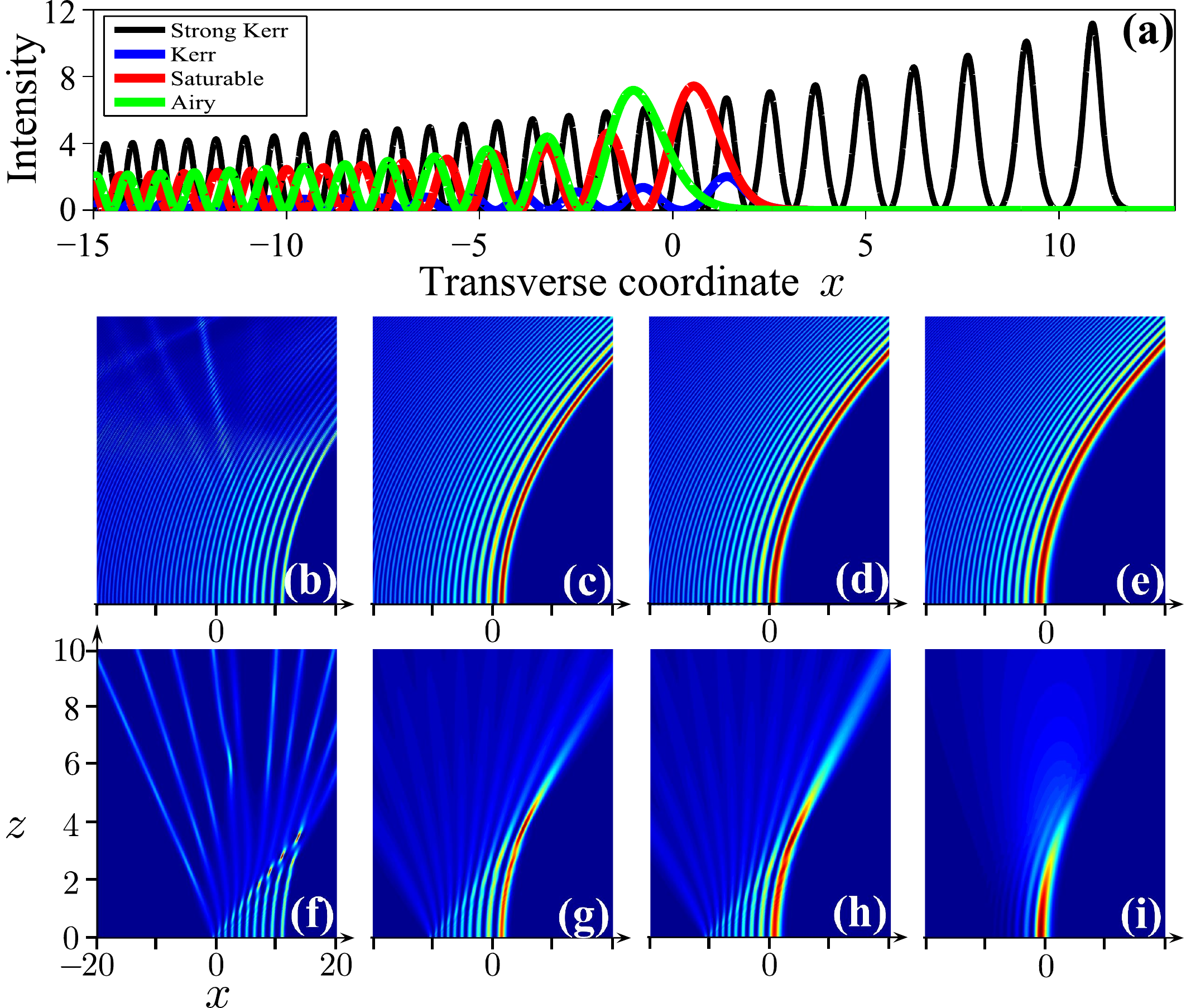}
  \caption{(a) Intensities of an Airy beam and nonlinear accelerating solutions at $z=0$.
  (b)-(e) Propagations of the strong Kerr, the Kerr, the saturable accelerating solution, and the Airy beam as shown in (a), respectively.
  (f)-(h) Propagation of the truncated nonlinear accelerating solutions that correspond to (b)-(d). Note the tendency of the
  shed radiation to move along the straight lines. (i) Propagation of a finite Airy beam that corresponds to (e).}
  \label{Fig1}
\end{figure}

To set the stage, in Fig. \ref{Fig1}(a)
we compare the intensities of the linear Airy function $5{\rm Ai}(x)$ and various nonlinear accelerating solutions at $z=0$.
The strongly NL Kerr solution is obtained for $\alpha=10^{12}$,
while the ``normal'' Kerr and saturable cases are obtained for $\alpha=30$.
The propagation properties of the four beams shown in Fig. \ref{Fig1}(a)
are displayed in Figs. \ref{Fig1}(b)-\ref{Fig1}(e),
from which one can see that the beams accelerate along a parabolic (that is, a square-root) trajectory.
In comparison with the linear Airy beam,
the main lobes of the nonlinear solutions accelerate more strongly in the positive transverse direction.
As is well known, the beams with high intensity may not be stable in the focusing Kerr media,
so the propagation of the strong Kerr accelerating beams is not stable, as is visible in Fig. \ref{Fig1}(b) and
\ref{Fig1}(f) -- some energy is shed from the lobes during propagation.

For the nonlinear accelerating beams, to obtain finite-energy beams we truncate them directly after 10 lobes,
at the appropriate value of $x$;
their propagation properties are shown in Figs. \ref{Fig1}(f)-\ref{Fig1}(h),
which correspond to Figs. \ref{Fig1}(b)-\ref{Fig1}(d), respectively.
Corresponding to Fig.  \ref{Fig1}(e),
the propagation of $5{\rm Ai}(x)\exp(ax)$ is shown in Fig. \ref{Fig1}(i).
It is seen that the main lobes can stably accelerate for quite a long propagation distance.
It is also evident that the individual solitons can be produced from the
main lobe of the truncated strong Kerr nonlinear solution during propagation,
while the other two NL solutions cannot generate solitons.
However, all the shed radiation and soliton-like beams tend to propagate along the straight lines.
Thus, the solitons, once formed, do not accelerate, but travel along straight trajectories.

Now that we have obtained the single-beam solutions of Eq. (\ref{eq1}) and
Eq. (\ref{eq4}), we want to investigate their interactions.
To this end, we construct more complex incident beams, made up of two shifted single beams, launched in parallel
but accelerating in opposite directions.
We will first investigate the interactions of two Airy beams,
so the incident beam will be composed of two shifted linear Airy solutions with a fixed relative phase between them,
\begin{align}
\label{eq5}  \psi(x)=  A_1 {\rm Ai} [ (x - B)] \exp[ a(x - B)] +
 \exp(il\pi) A_2 {\rm Ai} [- (x + B)] \exp[- a(x + B)].
\end{align}
Here $B$ is the transverse position shift and $l$ controls the phase shift.
If $l=0$, the two components are in-phase; if $l=1$, they are out-of-phase.
Here, we restrict our attention to these two values.
Also, we take $a=0.2$ throughout.

In addition, we will study the interactions of two NL solutions.
Because the nonlinear accelerating beams also possess infinite oscillating tails, as shown in Fig. \ref{Fig1}(a),
we will truncate them at a certain $x$ to keep only a certain number of lobes.
The incident beam is written as
\begin{align}
\label{eq7}  \psi(x)=  \psi_1 [ (x - B)] + \exp(il\pi) \psi_2 [- (x + B)],
\end{align}
where $\psi_1$ and $\psi_2$ are the two truncated NL solutions.

\section{Interactions of Airy beams}\label{airyairy}
\subsection{Linear medium}

\begin{figure*}[htbp]
  \centering
  \includegraphics[width=\textwidth]{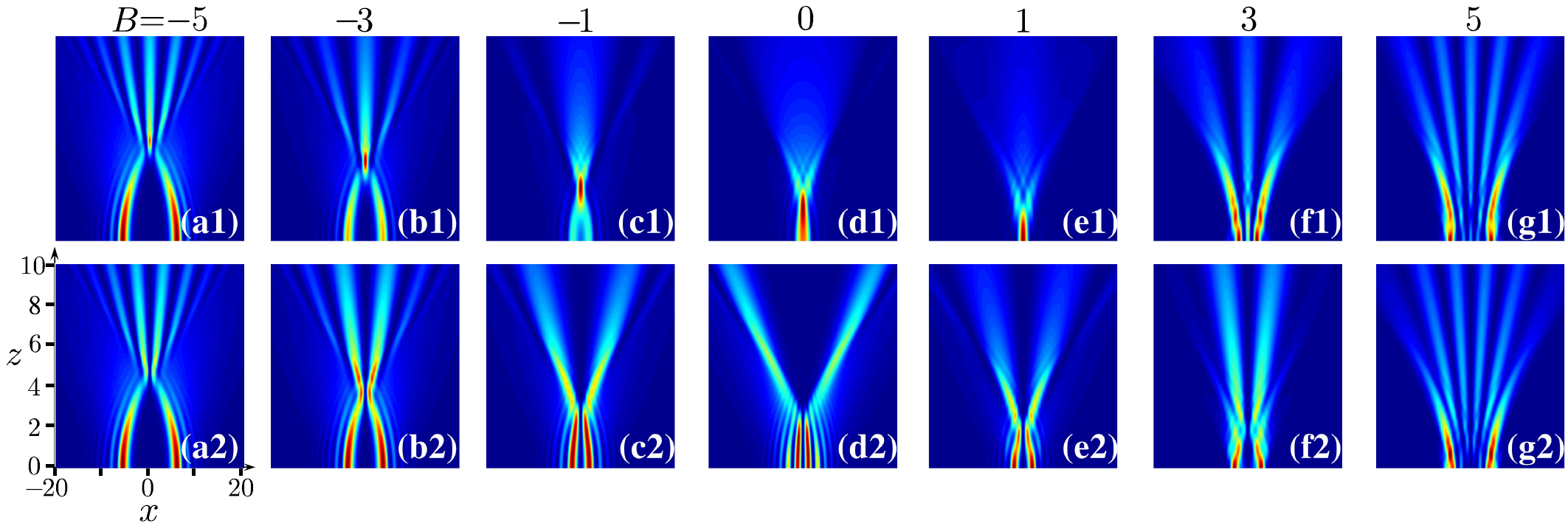}
  \caption{Interference of two incident Airy beams in the linear medium, with $A_1=A_2=4$ and different $B$.
  (a1)-(g1) In-phase case.
  (a2)-(g2) Out-of-phase case.}
  \label{Fig2}
\end{figure*}

Beam interaction takes place in NL media,
so in a linear medium, with $\delta n=0$ in Eq. (\ref{eq1}),
the ``interaction'' is actually a linear interference.
We display the evolution of the incidence from Eq. (\ref{eq5}), for different $B$;
the intensities are shown in Fig. \ref{Fig2},
in which the in-phase and out-of-phase cases correspond to $l=0$ and $l=1$, respectively.
The behavior shown in Figs. \ref{Fig2}(a1)-\ref{Fig2}(g1) and \ref{Fig2}(a2)-\ref{Fig2}(g2) is quite similar,
the major difference being that the central interference fringe in the in-phase case is bright, whereas in the out-of-phase case it is dark,
as it should be for a constructive and destructive interference.
It is seen that some mutual-focusing is observed in the central region, as the beams get closer.
It is caused by the interference of the curved accelerating beams.
The two Airy beams behave similar to that in Fig. 2(a) of Ref. \cite{siviloglou_ol_2007}.
They diffract, superpose, and interfere as they propagate. Of course, no solitons can form.

For an Airy beam, the energy is mainly stored in the main lobe; hence,
if the interval between the two main lobes of the incidence is large, there will be no mutual-focusing in the central region.
If the interval is small enough, the overlap of the main lobes will be considerable, leading to an apparent mutual-focusing.
This feature develops differently for the in-phase and out-of-phase beams;
in the first case, it happens in the central bright interference fringe,
while in the second case, it happens in the two first maxima.


\subsection{Kerr medium}

\begin{figure*}[htbp]
  \centering
  \includegraphics[width=\textwidth]{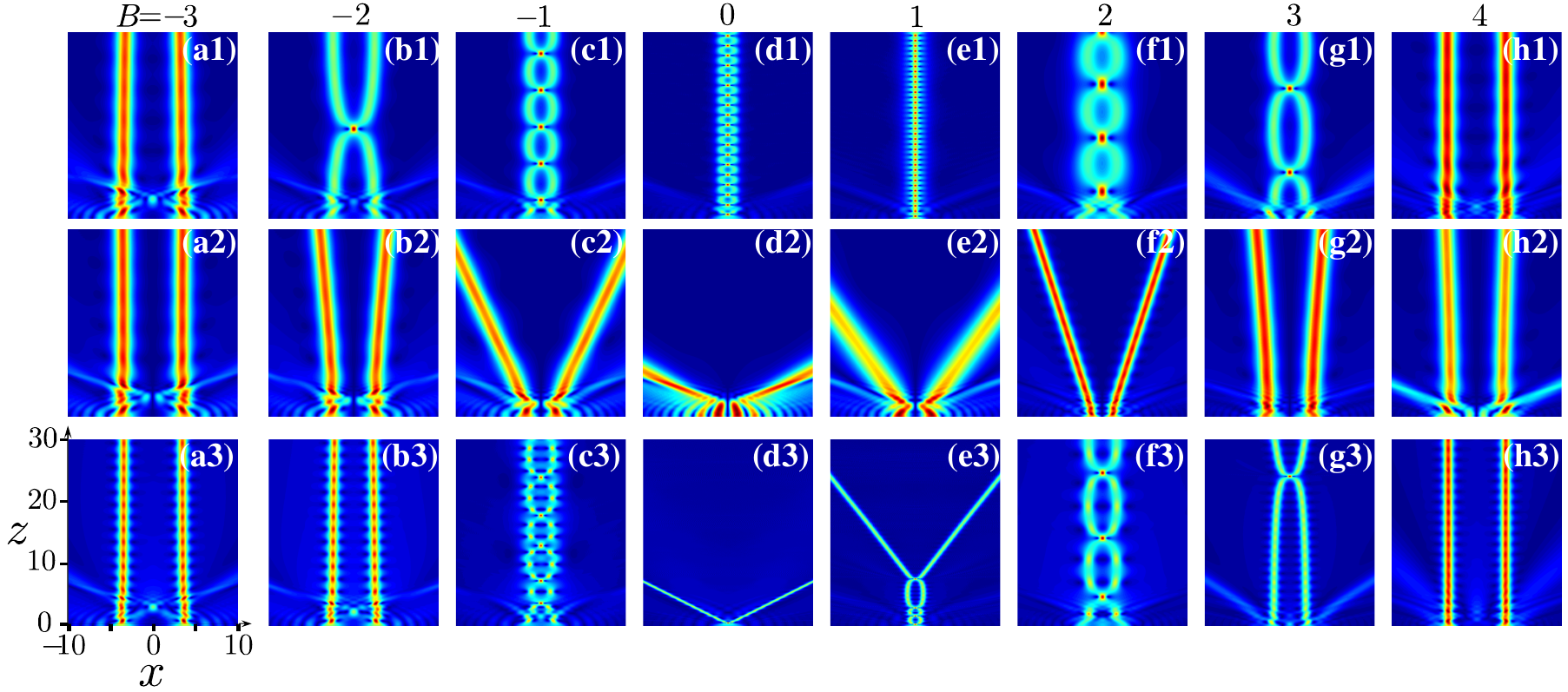}
  \caption{Soliton formation in the interaction of two in-phase ((a1)-(h1)) and out-of-phase ((a2)-(h2))
  incident Airy beams with $A_1=A_2=3$, in the Kerr medium.
  (a3)-(h3) The same as (a1)-(h1), but with $A_1=A_2=4$.
  }
\label{Fig3}
\end{figure*}

When $\delta n =|\psi(x)|^2$,
the nonlinearity is of the focusing Kerr-type.
Since a large interval between Airy components in the incidence leads to a weak interaction,
we just show the results with a relatively small interval.
We display the interactions of two in-phase and out-of-phase counter-accelerating Airy beams
with $A_1=A_2=3$ in Figs. \ref{Fig3}(a1)-\ref{Fig3}(h1) and \ref{Fig3}(a2)-\ref{Fig3}(h2), respectively.
Immediately visible is the considerable interaction and NL self- and mutual-focusing of the beams.
The major difference between the two cases is the attraction of beams in the in-phase case and the repulsion in the out-of-phase case.
Also visible is the breathing or filamentation of the beams when they strongly interact.
For $B=-3$ and 4 in the in-phase case, the two Airy components form two parallel solitons, after shedding some radiation,
as depicted in Figs. \ref{Fig3}(a1) and \ref{Fig3}(h1).
With the decreasing interval, the attraction between the two components increases
and bound breathing solitons are formed, with certain periods, as shown in Figs. \ref{Fig3}(b1)-\ref{Fig3}(g1).
The smaller the interval, the stronger the attraction and the smaller the period of soliton breathing.
Curiously, the intensity image shown in Fig. \ref{Fig3}(e1) has a smaller period than that in Fig. \ref{Fig3}(d1),
even though $B=0$ in that case.
A smaller interval between beams should produce larger interaction, which should lead to a smaller period.
The reason is that the main lobe of the Airy beam with $B=0$ is located at about $-1$,
and there is still an interval between the two main lobes in the incidence.
So, the attraction is the biggest when $B=1$ and the period of the formed soliton is then the smallest.
It is also worth mentioning that the solitons are generated from the main lobes and
that the acceleration property of the main lobes, as well as of the solitons,
is absent \cite{fattal_oe_2011,kaminer_prl_2011}. No accelerating beams as a result of interaction
are seen (unless one considers the wiggling breathers as ``accelerating'' beams.)

The results for the out-of-phase case are shown in Figs. \ref{Fig3}(a2)-\ref{Fig3}(h2),
which share the same numerical parameters as the in-phase case.
From the intensity images
one can see that the soliton pairs formed from the incidence actually repel each other.
The smaller the interval, the stronger the repulsion, until the beams start overlapping.
However, when the beams strongly overlap, like in Fig. \ref{Fig3}(e2), the repulsion decreases as the overlap increases.
Considering that the two Airy components are out-of-phase,
the main lobes will balance each other out at $B=1$,
so that the distance between the secondary lobes (Fig. \ref{Fig3}(e2))
is larger than the distance between the main lobes for $B=0$ (Fig. \ref{Fig3}(d2)).
In other words, the soliton pair shown in Fig. \ref{Fig3}(e2) is generated from the secondary lobes,
while the other are generated from the main lobes.
This is why the repulsion of the soliton pair in Fig. \ref{Fig3}(d2) is stronger than that in Fig. \ref{Fig3}(e2).

It is interesting to note that in Fig. \ref{Fig3}(h2), two soliton pairs are visible:
one pair comes from the main lobes of the Airy components and the other from the secondary lobes.
In all other figures only one pair is visible, in addition to the
excess radiation emanating initially from the interacting Airy beams.
Because the energy is mainly stored in the main lobes,
the intensity of the inner soliton pair is smaller than that of the outer pair,
but early in the propagation the two soliton pairs exchange energy at about $z=2$.
In addition, the repulsion of the outer soliton pair is stronger,
which comes from the main lobes possessing more energy.
We should note that these results will be different if $A$ is allowed to vary.
For small $A$ (less than 1), there will be no solitons generated;
for large $A$ ($\sim10$), multiple soliton pairs will be produced,
but the propagation may become unstable because of the strong self-focusing effect.

When one sets $A_1=A_2=4$, the corresponding results are shown in Figs. \ref{Fig3}(a3)-\ref{Fig3}(h3)
(only the in-phase cases are shown here, because the out-of-phase cases are similar to those with $A_1=A_2=3$,
shown in Figs. \ref{Fig3}(a2)-\ref{Fig3}(h2)).
Interestingly, we find that the repulsion between the two solitons may appear in the in-phase case,
especially for the cases $B=0$ and  $B=1$, as shown in Figs. \ref{Fig3}(d3) and \ref{Fig3}(e3).
When $B=1$,
the intensity of the superposed main lobes is enhanced, while the width is
suppressed, in comparison with the case in Fig. \ref{Fig3}(d3).
Thus, the two solitons generated from the splitting of the superposed main
lobes will experience a smaller repulsion force than that of Fig. \ref{Fig3}(d3).
The refractive index change will make the solitons attract each other,
and the attraction is quite strong over a long distance,
but eventually the repulsion will overtake the attraction, as shown in Fig. \ref{Fig3}(e3).
When $B$ is further increased, the main lobe of one component will superpose with the high-order lobes of the other component,
so the solitons come from the superposed main and high-order lobes, as shown in Fig. \ref{Fig3}(h3).
When the interval between the two solitons is large, the interaction between them becomes weak,
and they propagate parallelly, as in Figs. \ref{Fig3}(a) and \ref{Fig3}(h).

For the case when the two interacting Airy beams are of different amplitudes,
the energy distribution about $x=0$ will be asymmetric,
so the generated solitons will be of different intensities, propagating at different angles
or breathing asymmetrically (not shown).

\begin{figure}[htbp]
  \centering
  \includegraphics[width=\columnwidth]{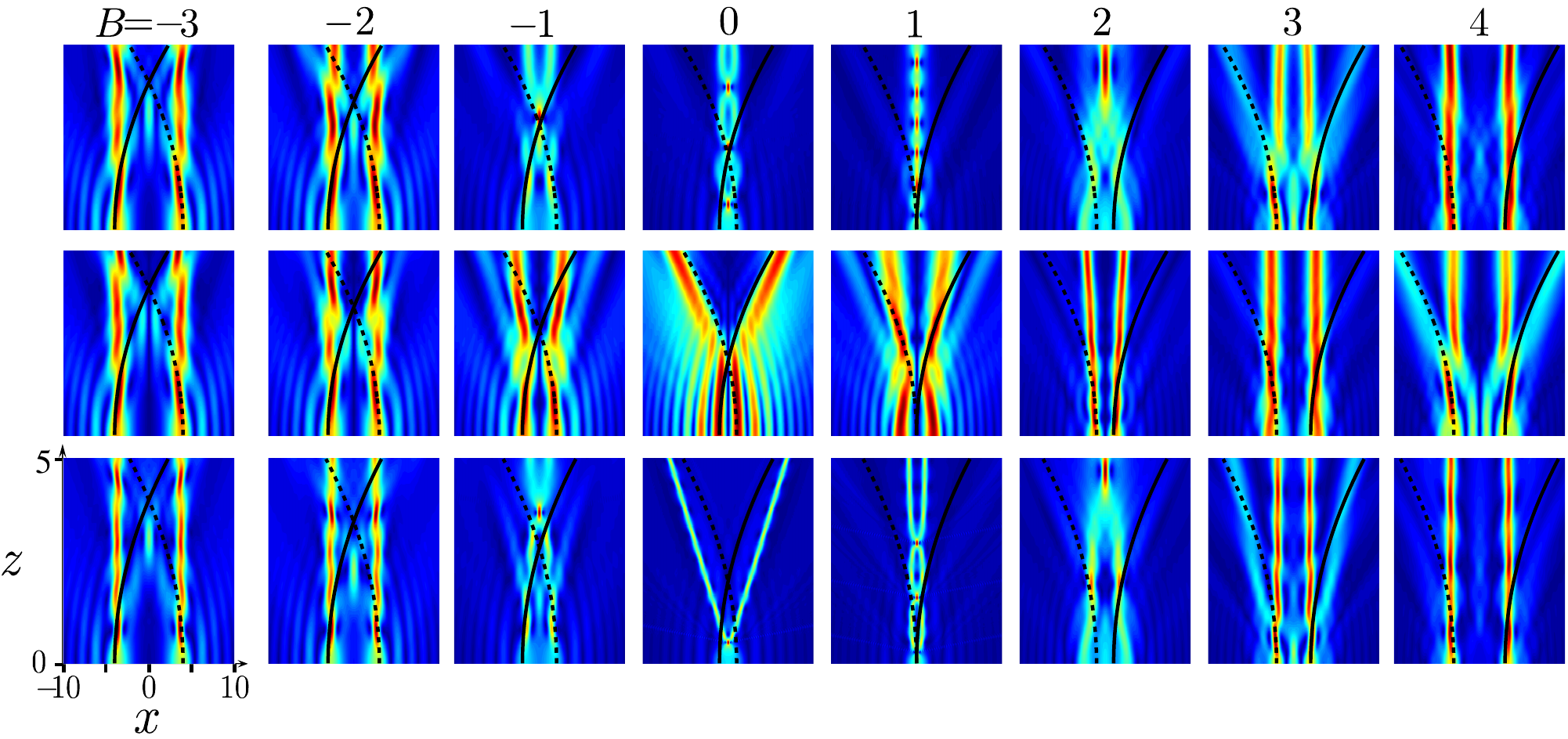}
  \caption{Same as Fig. \ref{Fig3}, but for shorter propagation distance.
The black solid and dashed curves present the ideal accelerating trajectories of the main lobes of the input Airy beams.}
\label{Fig3_short}
\end{figure}

To accentuate more clearly the generation of solitons from the lobes of Airy beams as they interact,
we replot Fig. \ref{Fig3}, but for shorter propagation distance and with an overlay
of the ideal accelerating trajectories of the main lobes of the Airy beams (Fig. \ref{Fig3_short}).
One can see that the accelerating property of the main lobes is gone; the solitons generated (or even the
plain radiation) are emitted tangentially from the main lobes.
The solitons, as well as the radiation shed, tend to travel along straight lines. Their interaction
during subsequent propagation also depends on the amplitude and the phase of the incident beams.
This would answer some of the questions raised in the introduction, whether there are accelerating beams emerging from
the interactions and whether the interaction is elastic. The short answers would be ``no'', and
the interaction is ``plastic''.

\subsection{Saturable medium}\label{airy-saturable}

\begin{figure*}[htbp]
  \centering
  \includegraphics[width=\textwidth]{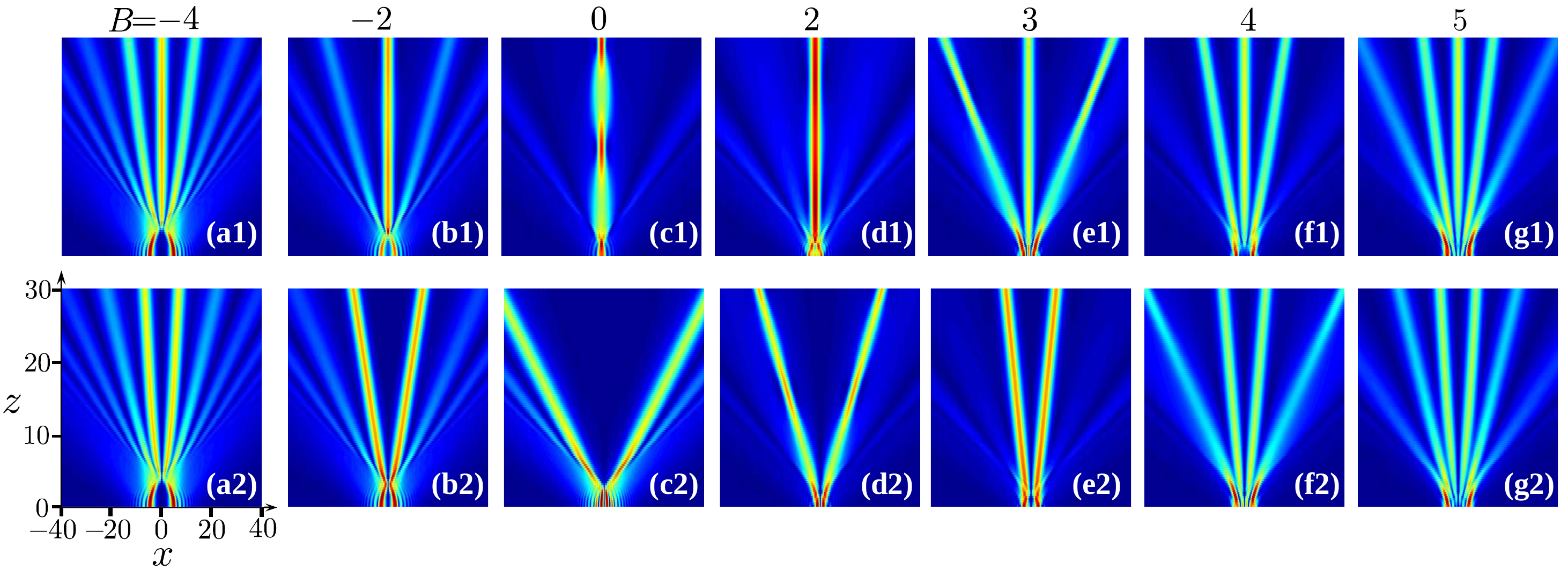}
  \caption{Interactions of two in-phase ((a1)-(g1)) and out-of-phase ((a2)-(g2)) Airy beams with $A_1=A_2=3$ in the saturable medium.
  }
\label{Fig6}
\end{figure*}

As mentioned above,
the saturable nonlinearity is of the form $\delta n =|\psi|^2/(1+|\psi|^2)$.
The numerical results are shown in Fig. \ref{Fig6}. Similar to the cases of Kerr medium,
the interactions can generate individual solitons and the soliton pairs with breather-like behavior.
As a rule, the in-phase cases generate the central individual solitons, whereas the out-of-phase cases do not.
For small $A_1$ and $A_2$, the solitons or soliton pairs cannot form in the interaction.
Importantly, the propagation in a saturable NL medium is stable for arbitrary $A_1$ and $A_2$, which is different from the Kerr medium.
From the numerical results in Fig. \ref{Fig6},
one can infer that the interactions in a saturable NL medium are also plastic,
similar to the cases in a Kerr NL medium.

In the end, we should mention that both Kerr and saturable NL media can support accelerating solutions, which possess infinite energy
-- but not accelerating solitons. In the manuscript, we only use the truncated beams to study the interactions.
Concerning interactions, there is nothing special about the saturable NL medium.
The main difference between the two types of nonlinear media is the stability at higher intensities.
The accelerating beams cannot behave like particles or solitons -- they are only extended beams,
even when propagating in the NL media. After the interaction, solitons or solitons pairs are generated from the colliding main lobes,
as well as radiation.
The generated solitons are just the commonly known solitons, which exhibit particle properties and do not accelerate.


\section{Interactions of nonlinear accelerating beams}\label{nonnon}

We now address the interactions of nonlinear accelerating beams.
In Fig. \ref{Fig1} we have displayed numerically obtained Kerr, saturable, and strong Kerr nonlinear accelerating modes.
Similar to the Airy modes, they exhibit long tails and possess infinite energy.
As seen in Figs. \ref{Fig1}(b)-\ref{Fig1}(e), the Airy beam as well as the nonlinear beams accelerate
along parabolic (or square-root) trajectories.
Because of the infinite power,
it is reasonable to cut off oscillating tails and study the truncated cases, shown in Figs. \ref{Fig1}(f) and \ref{Fig1}(h).
One can see that the truncated ``normal'' Kerr solutions (obtained for $\alpha=30$) as shown in Fig. \ref{Fig1}(g), shed radiation but
form no solitons. However, the strong Kerr solutions (obtained for $\alpha=10^{12}$), shown in Fig. \ref{Fig1}(f),
readily form solitons from the strong radiation shedding.

\subsection{Kerr medium}\label{nonkerr}

\begin{figure}[htbp]
  \centering
  \includegraphics[width=\columnwidth]{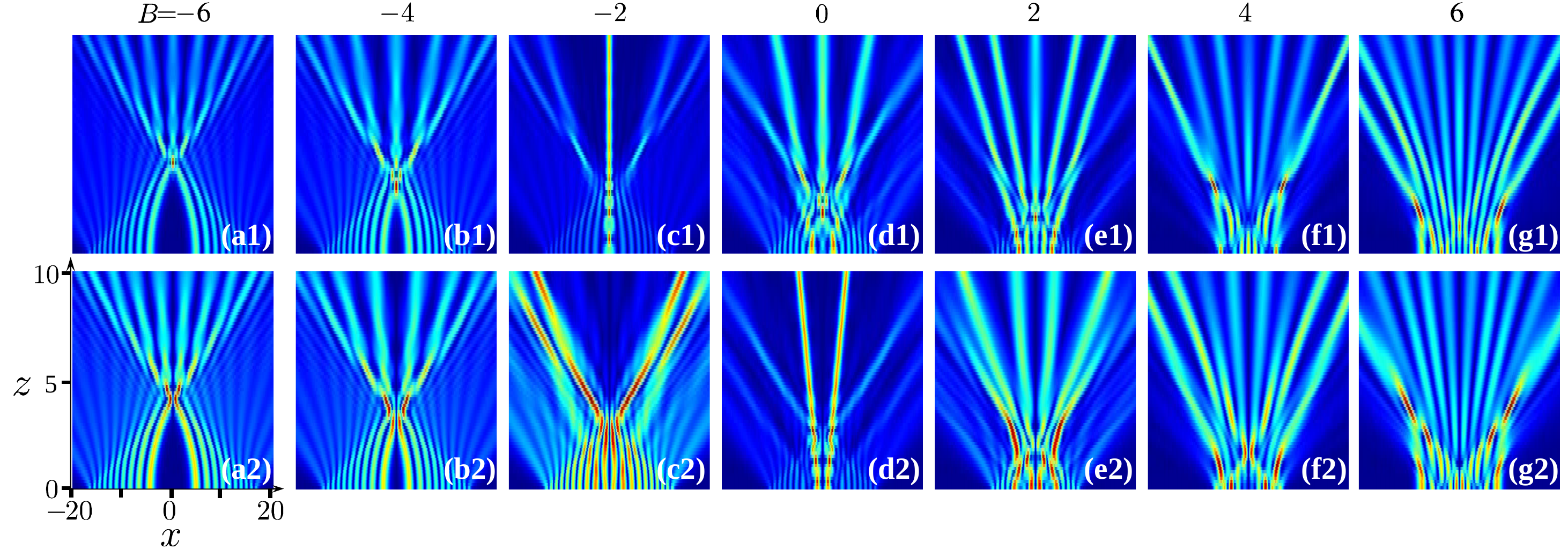}
  \caption{Interactions of two in-phase (top panels) and out-of-phase (bottom panels)
  nonlinear Kerr accelerating beams with $\alpha=30$, respectively.}
\label{Fig7}
\end{figure}

Similar to the interactions of Airy beams, we now study the interactions
of two truncated nonlinear beams in NL media, according to Eq. (\ref{eq7}).
Figure \ref{Fig7} depicts the interactions of two in-phase (top panels) and out-of-phase
(bottom panels) truncated Kerr beams obtained with $\alpha=30$ during acceleration.
It is seen that solitons and solitons pairs can be produced in the interactions.
Also, before the formation of soliton pairs, interactions will lead to extreme focusing of the beams
(red dots in the figures), which decreases the brightness in the panels of Fig. \ref{Fig7}.
In general, one obtains complex beam patterns.
With the increasing interval between the two components,
the interaction becomes weaker gradually, as seen in Figs. \ref{Fig7}(a), \ref{Fig7}(b), \ref{Fig7}(f) and \ref{Fig7}(g),
from which one can also see that the main lobes of the nonlinear accelerating beams still exhibit some accelerating property,
because they do not interact with each other over a long distance.
When the two nonlinear accelerating beams overlap more and interact with each other more strongly,
the main lobes do not show the accelerating property any longer, as shown in Figs. \ref{Fig7}(d)-\ref{Fig7}(e),
because most of the energy of the main lobes contributes to the formation of the solitons.
Thus, the interactions are also plastic.

When the truncated strong Kerr solutions interact, complex beam patterns are seen,
including individual solitons, breathing solitons, and radiation (not shown).
It is found that the accelerating beams easily generate single solitons and soliton pairs, which do not accelerate.
In comparison with the normal Kerr solutions, more solitons and soliton pairs can be induced from the strong Kerr solutions.
Similar to the cases in Fig. \ref{Fig7}, the interactions are also plastic.

%
%

\subsection{Saturable medium}
As mentioned in Sec. \ref{airy-saturable},
the behavior of the saturable nonlinear accelerating modes is quite similar to that of the Kerr nonlinear modes,
except that the saturable NL medium endorses stable propagation of beams with arbitrary high intensities.
Therefore, we do not discuss them here.

\section{Interactions of different accelerating beams}\label{different}

Up to now we have considered interactions of like beams: two Airy beams, two accelerating nonlinear beams, etc.
However, one can consider the interaction of unlike beams; then
many possibilities arise.
As an example, we will cover a case of single soliton interacting with
a Kerr accelerating beam. This case is interesting because a narrow solitary beams interacts with
a wide accelerating beam. In this case the accelerating beam will tend to retain its accelerating
property and the interaction will be more elastic, with the exclusion of the specific situation when
the soliton strongly overlaps with the main lobe of the accelerating beam.

\begin{figure}[htbp]
  \centering
  \includegraphics[width=\columnwidth]{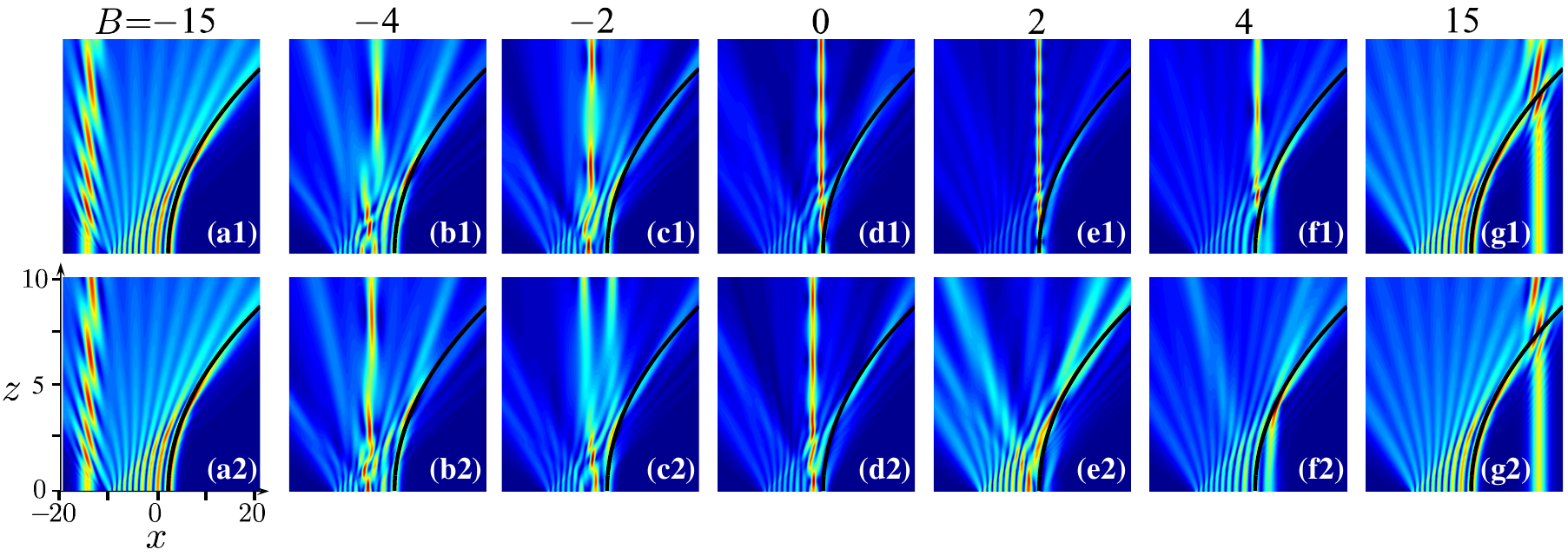}
  \caption{Interactions of in-phase (top row) and out-of-phase (bottom row) soliton and truncated Kerr accelerating beams.
  Black curves show the accelerating trajectories of the main lobe.}
\label{Fig12}
\end{figure}

As is well known, Eq. (\ref{eq1}) has a stationary soliton solution of the form
\begin{equation*}
  \psi(x,z)={\rm sech}(x)\exp(iz/2),
\end{equation*}
when $\delta n=|\psi|^2$.
For other nonlinearities (e.g., saturable nonlinearity), analytical soliton solution of Eq. (\ref{eq1}) cannot be found,
but one can use numerical methods to find a numerical solution \cite{yang_book}.
Here, we investigate the interaction between a soliton solution and a Kerr nonlinear accelerating solution,
by launching a composite beam $\psi(x)=  \psi_1 (x) +\exp(il\pi) \psi_2 [ (x - B)]$,
in which $\psi_1$ and $\psi_2$ represent the truncated Kerr nonlinear accelerating solution and the single soliton solution, respectively.
The corresponding numerical results are shown in Fig. \ref{Fig12}.

When the distance between the two components is big, as shown in Figs. \ref{Fig12}(a) and \ref{Fig12}(g),
the soliton will collide with the relatively weak lobes of the nonlinear accelerating beam; that is why
the intensity of the soliton exhibits fluctuations, while the Kerr beam still accelerates.
When the distance between components is smaller, as in Figs. \ref{Fig12}(b)-\ref{Fig12}(f),
the propagation properties depend on the profile of the superposed beam. In principle, the emerging breathing soliton
will come from the soliton component, modulated by the interaction with the lobes of the Kerr accelerating beam.
No soliton is generated from the main lobe.
As a consequence, the properties of the soliton and the nonlinear accelerating beam are quite immune to the collision,
owing to the conservation laws and the stability of both beams.
Comparing Fig. \ref{Fig12}(a1) with Fig. \ref{Fig12}(a2) (or Fig. \ref{Fig12}(g1) with Fig. \ref{Fig12}(g2)),
we can see that if the collision leads to a peak in the in-phase case,
it will lead to a dip in the out-of-phase case, during propagation.

In Figs. \ref{Fig12}(a) and \ref{Fig12}(g), the main lobes conserve the accelerating property, due to the same reason as in Sec. \ref{nonkerr}
-- a weak interaction with the soliton.
When the main lobe interacts with the soliton, it is more strongly affected both in amplitude and width.
Different from the strongly interacting cases of wide beams mentioned previously, in Fig. \ref{Fig12}
we can find some energy distributions along the curved black lines (even though there is also some energy shedding).
This can be attributed to the retention of the accelerating property, owing to the limited transverse extension of the single soliton.
Thus, we can say that the the main lobe still exhibits some accelerating property.

\section{Conclusion}\label{conclusion}
In summary, we have investigated the interactions of both in-phase and
out-of-phase Airy beams, nonlinear accelerating beams, and the soliton beams propagating in Kerr and saturable NL media.
We find that single solitons and soliton pairs can be produced in these interactions.
As the Airy beams and nonlinear accelerating beams have infinite energy,
we use the corresponding truncated beams, to discuss their interactions.
The generated individual solitons and soliton pairs do not accelerate transversely,
because their properties are determined by the
underlying NL media and not by the incident beam from which they are generated. They fly straight away along the
tangential lines to the lobes from which they are generated.
However, the frequency of generation as well as the subsequent interactions between solitons do depend on
the amplitude and phase of the incident beams.
The interactions of Airy beams and nonlinear accelerating beams are plastic, in
that the main lobes of the beams do not show accelerating property after the interaction.
Most of the energy of the main lobes participates in the formation of solitons or soliton pairs during the plastic interaction.
If an accelerating beam interacts with a single soliton,
the accelerating property of the main lobe of the accelerating beam is still retained after interaction, owing to the limited extension of
the soliton and insufficient interaction to produce solitons from the main lobe.

\section*{Acknowledgement}
This work was supported by
the China Postdoctoral Science Foundation (No. 2012M521773) and
the National Natural Science Foundation of China (No. 61308015).
Work in Qatar is supported by the NPRP 09-462-1-074 project from the Qatar National Research Fund (a member of the Qatar Foundation).
The following programs in PR China are also acknowledged: 973 Program (No. 2012CB921804),
the National Natural Science Foundation of China (Nos. 11104214, 61108017, 11104216, 61205112),
the Specialized Research Fund for the Doctoral Program of Higher Education of China (Nos. 20110201110006, 20110201120005),
and the Fundamental Research Funds for the Central Universities (Nos. xjj2013089, 2012jdhz05, 2011jdhz07, xjj2011083, xjj2011084, xjj2012080).

\end{document}